\documentclass[twocolumn,showpacs,amsfonts,amsmath,amssymb,prb]{revtex4}
\usepackage{epsfig}
\usepackage{subfigure}
\usepackage{amsmath}
\usepackage{amssymb}
\usepackage{array}
\usepackage{pifont}
\usepackage{color}
\definecolor{grey}{rgb}{0.4,0.4,0.4}
\newcommand{\grey}{\color{grey}}
\begin{document}
\title{Simple model for a quantum wire II. Correlations}%
\author{Jose M. Cerver\'o}%
  % \email{cervero@sonia.usal.es} 
   \email[To whom correspondence should be addressed:\\]{cervero@sonia.usal.es}%
    \homepage[\\Visit: ]{http://sonia.usal.es/qwires/}
\author{Alberto Rodr\'{\i}guez}%
    \affiliation{F\'{\i}sica Te\'orica. Facultad de %
Ciencias. Universidad de Salamanca. 37008 Salamanca. Spain}%
\begin{abstract}%
  In a previous paper (\textbf{Eur. Phys. J. B 30}, 239-251
(2002)) we have presented the main features and properties of a simple
model which -in spite of its simplicity- describes quite accurately the
qualitative behaviour of a quantum wire. The model was composed of N
distinct deltas each one carrying a different coupling. We were able to
diagonalize the Hamiltonian in the periodic case and yield a complete and
analytic description of the subsequent band structure. Furthermore the
random case was also analyzed and we were able to describe Anderson
localization and fractal structure of the conductance. In the present paper
we go one step further and show how to introduce correlations among the
sites of the wire. The presence of a correlated disorder manifests itself
by altering the distribution of
states and the localization of the electrons within the system
\end{abstract}
\pacs{03.65.-w: Quantum Mechanics, 71.23.An: Theories and Models;
Localized States, 73.21.Hb: Quantum Wires}
\maketitle
\section*{Introduction}
\label{sec:intro}
    In a previous paper \cite{QWI}, hereafter referred to as I, the authors
have developed a simple model describing the main features shown by a
Quantum Wire, namely: Band Structure when the structure is periodically
arranged, Anderson Localization in the random case and easy calculation of
Lyapunov exponents, density of electronic states and electrical conductance
endowed with an encouraging fractal behaviour. We address the interested
reader to reference I for the details of the model and the intrincacies of
the computer calculations.

The main drawback of the scheme thereby presented was the lack of a
coherent description of the possible correlations which must be present in
any realistic model of one dimensional conductance. It is the purpose of
the present paper to fill this gap by presenting a surprisingly simple way to
incorporate correlations entirely based upon the random nature of the
model. This means that in addition to describe correlations with different
parameters of the wire (spacing, number and amount of species, ... ) the
model is flexible enough to be suitable for applying it to different
correlation schemes such as the Tight-Binding model, the Anderson model and
likely many others correlation prescriptions with sensible physical
content.

The paper is organized as follows. In Section 1 we analyze the meaning of
the correlations in terms of probabilities of a given atom to be followed
by other of a different character. After this analysis we are able to set
some limits for the values of these probabilities. The next step is to
incorporate those probabilities to the Functional Equation developed in
I. How this functional equation is modified by the introduction of
correlations in the way just described and how it can be used to calculate 
the correlation-corrected Lyapunov exponent is the subject of Section
2. In Section 3 some results are shown along with a discussion about 
the effects of the correlations to close finally with a section
of Conclusions.
%%%%%%%%%%%%%%%%%%%%%%%%%%%%%%%%%%%%%%%%%%%%%%%%%%%%%%%%%%%%%%%%%%%
\section{Correlations}
\label{sec:corre}
Let us consider disordered quantum wires with short-range
correlations which will be specified by the probability for a certain
atomic species to be followed by another type of atom in the chain
sequence. In this model, the system will be characterized by the species
concentrations $\{c_i\}_{i=1,\ldots,m}$ where $m$ is the number of
different species, and the set $\{p_{ij}\}_{i,j=1,\ldots,m}$ where
$p_{ij}$ means the probability for $i$ to be followed by $j$ (the
prob. of finding $j$ right after $i$). The correlations introduced in this way
 must be the consequence of the existence of an atomic interaction (or the
effect of some physical parameters such as the size of the atoms) which
might choose certain spatial sequences of the atoms modifying subtly the 
otherwise chain's purely random character. Thus this procedure seems a natural manner
to account for correlations which can be present in nature or even those that
can be produced inside a manufactured disordered quantum wire in a nonintentional
way aside from the fact that of course one can always try to construct a
wire exhibiting the desired correlations to recover the theoretical results.

Let us clarify the meaning of $p_{ij}$. It is the ratio of the number of
-$ij$- clusters  and the number of $i$ atoms ($N_i$). The probability
$p_{ij}$ is not the probability for $j$ to be preceded by $i$, because the
latter would be obtained dividing by the number of $j$ atoms ($N_j$) instead
of $N_i$. In the limit $N\rightarrow\infty$, the system will be determined
by $\{c_i\}$ and $\{p_{ij}\}$ and therefore the density of states will be
given by a limiting distribution depending only on those quantities. And
that distribution is of course the same whether we move from left to
right or from right to left along the chain. So, the correlations we measure
have to coincide in both directions. Then it is clear that \textit{for the infinite chain
 $p_{ij}$ means the probability for an '$i$' species atom to be followed or
preceded by a '$j$' species atom.}

Once we have understood the meaning of the correlations, let us derive the
relations among them. What is the probability of finding at any position of
the chain the cluster -iji-? We can write this quantity as,
\begin{align*}
    c_i\cdot p_{ij}\cdot p_{ji} \rightarrow & \text{\small prob. of finding $i\,\cdot$ 
    finding $j$ after $i\,\cdot$}\\
    & \text{\small $\cdot$  finding $i$ after $j$}\\
    c_j\cdot p_{ji}^2 \rightarrow & \text{\small prob. of finding $j\,\cdot$ finding
    $i$ after $j\,\cdot$}\\
    & \text{\small $\cdot$ finding $i$ before $j$}.
\end{align*}
Thus the equations for the correlations are,
\begin{subequations}
\label{eq:corre}
\begin{align}
    c_i p_{ij} = c_j p_{ji} &\label{eq:corre_a}\\
    \sum_{j=1}^m p_{ij} = 1  & \label{eq:corre_b} \\
    0\leq p_{ij}  \leq 1 & \quad i,j=1,\ldots,m \label{eq:corre_c}.
\end{align}
\end{subequations}
According to the equations, the correlations matrix $(p_{ij})$ is completely known
from the above diagonal elements, $\frac{m(m-1)}{2}$ where $m$ is the
number of different species. However these elements are not completely
independent due to \eqref{eq:corre_c}, because when one of the correlations is
chosen, the maximum allowed value for some of the rest may be affected.
This fact can be clearly seen in two simple examples:
\begin{description}
\item[\bf 2 species:]
    the matrix is completely determined by $p_{12}$,
    \begin{equation*}
        p_{11} = 1-p_{12}\quad;\quad
        p_{21} = \frac{c_1}{c_2} p_{12}\quad;\quad
        p_{22} = 1-p_{21}
    \end{equation*}
    but it must be $p_{21}\leq1 \Rightarrow p_{12}\leq\dfrac{c_2}{c_1}$.

    Therefore \mbox{$p_{12}\leq min\left\{1,\dfrac{c_2}{c_1}\right\}$}.
    
\item[\bf 3 species:]
    the matrix is determined by $p_{12}, p_{13}, p_{23}$. And we choose
their values in that order.

 \mbox{$p_{12}\leq min\left\{1,\dfrac{c_2}{c_1}\right\}$} which implies 
   
\mbox{$p_{13}\leq min\left\{1-p_{12},\dfrac{c_3}{c_1}\right\}$}. 

On the other hand it must be,
\begin{gather*}
     p_{23}\leq 1-p_{21}\\
     p_{32}\leq 1-p_{31} \Rightarrow p_{23}\leq \frac{c_3}{c_2}(1-p_{31}),
\end{gather*}
 therefore 
 \mbox{$p_{23}\leq min\left\{1-\dfrac{c_1}{c_2}p_{12},\dfrac{c_3}{c_2}-\dfrac{c_1}{c_2}p_{13}\right\}$}.
\end{description}    

An expression for the general form of the maximum values can be obtained
for arbitrary $m$. Choosing the above diagonal elements of the correlations
matrix by rows,
\begin{equation*}
\mathbf{P}=\begin{pmatrix}
     p_{11} &  \grey p_{12} & \grey \longrightarrow & \grey \ldots &\grey\longrightarrow & \grey p_{1m} \\
     p_{21} &  p_{22} & \grey  p_{23} &  \grey \longrightarrow& \grey\longrightarrow & \grey p_{2m} \\
     \vdots &  \vdots &  \ddots & \grey \ddots & & \grey \vdots \\
     \vdots &  \vdots & &  \ddots & \grey \ddots & \grey \vdots \\
     p_{(m-1)1} &  p_{(m-1)2} & \hdotsfor{2} &  p_{(m-1)(m-1)}  & \grey p_{(m-1)m}\\
     p_{m1} &  p_{m2} &  \hdotsfor{2}& p_{m(m-1)} & p_{mm}
    \end{pmatrix}
\end{equation*}
then 
\begin{equation}
    p_{ij}\leq min\left\{1-\frac{1}{c_i}\sum_{k=1}^{i-1}c_k
    p_{ki}-\sum_{k=i+1}^{j-1}p_{ik};\, \frac{c_j}{c_i}-\frac{1}{c_i}\sum_{k=1}^{i-1}c_k
    p_{kj}\right\}
\label{eq:limit}
\end{equation}
where the limits are given by the previous chosen correlations.

Once the concentrations and $\mathbf{P}$ are fixed, the
characterization of the quantum wire can be given in a compact form using the
matrix 
\begin{equation}
    \mathbf{Q}=(q_{ij})\equiv(c_i p_{ij}),
\end{equation}
which is symmetric ($\mathbf{Q}=\mathbf{Q}^t$) and it determines uniquely
the wire and its correlations:
\begin{equation*}
    \sum_{j=1}^m q_{ij}=c_i \quad;\quad \sum_{i=1}^m q_{ij}=c_j \quad;\quad
p_{ij}=\frac{q_{ij}}{c_i} \quad;
\end{equation*}
 $q_{ij}$ means the probability of finding the cluster -ij- (-ji-) at
any position inside the wire.

\begin{figure}
    \centering
    \epsfig{file=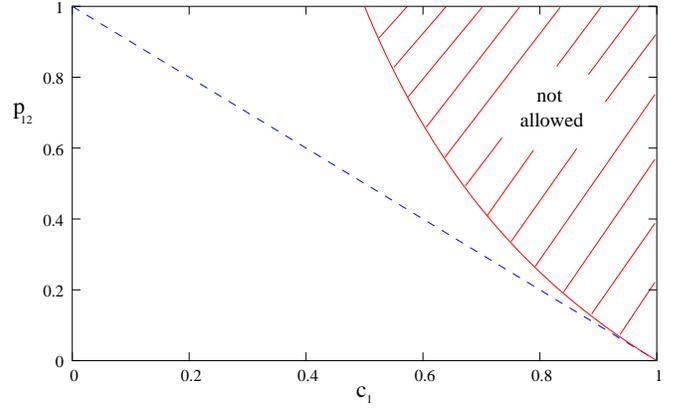,width=\columnwidth}
    \label{fig:space}
    \caption{Correlation space for 2 species as a function of the
concentration. The blue dashed line corresponds to the completely random configurations.}
\end{figure}

This model naturally includes the situation in which the
disorder in the wire is completely random\footnote{The case in which the wire is obtained ``throwing'' the atoms in the given
concentrations randomly and there are no mechanisms (interactions) which
select some privileged clusters or series of atoms.}, that is just defined
 by the values $p_{ij}=c_j \quad i=1,\ldots,m$. Figure \ref{fig:space}
shows an example of the correlation space for 2 species.
%%%%%%%%%%%%%%%%%%%%%%%%%%%%%%%%%%%%%%%%%%%%%%%%%%%%%%%%%%%%%%%%%%%%%%
\section{The functional equation}
\label{sec:FE}
    We will now obtain the correlation form of the
functional equation, which enable us to calculate the density of states. In
this process we are using some of the results and expressions obtained in reference I.

The relations for the phase transmission inside the wire read,
\begin{subequations}
\label{eq:Tinverse}
\begin{gather}
    \mathcal{T}_j^{-1}(\theta)=\arctan\left(
    2 h_j(\epsilon) -
    \frac{1}{\tan\theta}\right) \label{eq:Tinverse_a}\\
    \mathcal{T}_j^{-1}(\theta+n\pi)=\mathcal{T}_j^{-1}(\theta)+n\pi
    \qquad \theta \in\left[0,\pi\right).
\end{gather}
\end{subequations}
where $h_j(\epsilon)=\cos(\epsilon)+\frac{(a/a_j)}{\epsilon}\sin(\epsilon)$.
To carry out the phase average we rearrange the terms depending on the
species of the last atom we have passed,
\begin{multline}
    \langle\Delta\theta\rangle = c_1\frac{1}{N_1}\sum_{j_1}
    \left\{\mathcal{T}_{j}(\theta_{j})-\theta_{j}\right\}+ \ldots \\\ldots +c_m\frac{1}{N_m} \sum_{j_m}
    \left\{\mathcal{T}_{j}(\theta_{j})-\theta_{j}\right\}
\end{multline}
where $N_i$ is the number of atoms of species $i$ and the sumatory over
$j_i$ means we are summing over all the positions $j$ such that the atom at
$(j-1)$ is of $i$-kind:
\begin{equation}
   \langle\Delta\theta\rangle=c_1 \langle\Delta\theta\rangle_1
 + \ldots + c_m\langle\Delta\theta\rangle_m = \sum_\gamma c_\gamma
\langle\Delta\theta\rangle_\gamma.
\end{equation}  
From now on the greek indices will refer to species while the latin indices
mean the sites of the chain. In this case there exist correlations in the
sequence of atoms, that is the probability for one species to be a
neighbour of 
another type can be different for all of them. This implies that each
species will give rise to a particular distribution function for the
phase.
\begin{equation}
    \langle\Delta\theta\rangle_\gamma=\frac{1}{N_\gamma}\sum_{j_\gamma}\int_0^\pi 
    \frac{dW_j^\gamma(\theta)}{d\theta} \left\{\mathcal{T}_j(\theta)
-\theta\right\} d\theta
\end{equation}
where $W_j^\gamma(\theta)$ is the distribution function of the phase at the
position $j$ generated by the
species $\gamma$. The equations for the distributions ((B.14) in I) become here,
\begin{subequations}
\begin{gather}
     W_j^\gamma(\theta)=W_{j-1}^{(*)}\left(\mathcal{T}_\gamma^{-1}(\theta)\right) - 
    W_{j-1}^{(*)}\left(\frac{\pi}{2}\right)+1 \\ 
    W_j^\gamma(\theta+r\pi)=W_j^\gamma(\theta)+r \quad \theta\in[0,\pi) \quad r\in\mathbb{Z}
\end{gather}
\end{subequations}
for all species and positions, where the superscript ``$(*)$'' means that we do not
know the species of the atom preceding the $\gamma$ atom,
placed at position $(j-1)$; we only know the probability for a certain
species to be there.
Let us introduce now the average over all possible sequences with the given
concentrations.
\begin{equation}
    \langle\Delta\theta\rangle_\gamma=\frac{1}{N_\gamma}\sum_{j_\gamma}\int_0^\pi 
    \frac{dW_j^\gamma(\theta)}{d\theta} \sum_\beta p_{\gamma\beta}\left\{\mathcal{T}_\beta(\theta)
-\theta\right\} d\theta
\end{equation}
and,
\begin{equation}
    W_j^\gamma(\theta)=\sum_\beta p_{\gamma\beta} \left\{W_{j-1}^\beta \left(\mathcal{T}_\gamma^{-1}(\theta)\right) - 
    W_{j-1}^\beta\left(\frac{\pi}{2}\right)\right\}+1. 
\end{equation}
The next step consists in taking the limit $N\rightarrow\infty$ to approach the
limiting distribution. 

We define $\displaystyle\mathbf{W}_\gamma(\theta)=\lim_{N_\gamma\rightarrow\infty}\frac{1}{N_\gamma}
\sum_{j_\gamma}W^\gamma_j(\theta)$ which is the solution of:
\begin{subequations}
\label{eq:Wcorre}
\begin{gather}
     \mathbf{W}_\gamma(\theta)=\sum_\beta p_{\gamma\beta} \left\{\mathbf{W}_\beta \left(\mathcal{T}_\gamma^{-1}(\theta)\right) - 
    \mathbf{W}_\beta\left(\frac{\pi}{2}\right)\right\}+1 \label{eq:Wcorre_a}\\
    \mathbf{W}_\gamma(\theta+r\pi)=\mathbf{W}_\gamma(\theta)+r \quad
\theta\in[0,\pi) \quad r\in\mathbb{Z}. \label{eq:Wcorre_b}
\end{gather}
\end{subequations}
And finally we have to calculate,
\begin{equation}
    \langle\Delta\theta\rangle=\sum_{\gamma,\beta} c_\gamma
    p_{\gamma\beta}\int_0^\pi  \frac{d\mathbf{W}_\gamma(\theta)}{d\theta}\left\{\mathcal{T}_\beta(\theta)
    -\theta\right\} d\theta.
\end{equation} 
    To carry out the integral we shall make use of the existence of a value
$\theta_0$ such that $\mathcal{T}_\gamma(\theta_0)=\theta_1 \quad
\forall\,\gamma$, and proceeding by parts,
\begin{multline}
     \langle\Delta\theta\rangle = (\theta_1-\theta_0)
    -\overset{\text{\small\ding{172}}}{\overbrace{\sum_{\gamma,\beta}c_\gamma p_{\gamma\beta} \int_{\theta_0}^{\theta_0+\pi}
    \mathbf{W}_\gamma(\theta)\frac{d\mathcal{T}_\beta(\theta)}{d\theta} d\theta}}\\
    +\sum_\gamma c_\gamma \int_{\theta_0}^{\theta_0+\pi}
    \mathbf{W}_\gamma(\theta)d\theta.
    \label{eq:finalav} 
\end{multline}
 To evaluate the middle term we use the average of equations
\eqref{eq:Wcorre_a} which is
\begin{equation}
    \sum_\gamma c_\gamma \mathbf{W}_\gamma(\theta) +\sum_\beta c_\beta
    \mathbf{W}_\beta\left(\frac{\pi}{2}\right) -1 = \sum_{\gamma,\beta}
    c_\gamma p_{\gamma\beta}
    \mathbf{W}_\beta\left(\mathcal{T}_\gamma^{-1}(\theta)\right).
    \label{eq:Wcorreav}
\end{equation}
With a simple change of variable, the use of \eqref{eq:Wcorreav} and the
equations for the correlations we can
write,
\begin{multline}
    \text{\ding{172}}= \sum_{\gamma,\beta} c_\beta p_{\beta\gamma}
    \int_{\theta_1}^{\theta_1+\pi}
    \mathbf{W}_\gamma\left(\mathcal{T}^{-1}_\beta(\theta)\right) d\theta = \\=
     \sum_\gamma c_\gamma
    \int_{\theta_1}^{\theta_1+\pi}\mathbf{W}_\gamma(\theta) d\theta+ \pi \sum_\beta c_\beta
    \mathbf{W}_\beta\left(\frac{\pi}{2}\right) -\pi,
\end{multline}
 and going back to \eqref{eq:finalav},
\begin{multline}
    \langle\Delta\theta\rangle= (\theta_1-\theta_0) +\pi - \pi\sum_\beta c_\beta
    \mathbf{W}_\beta\left(\frac{\pi}{2}\right)+\\+\sum_\gamma c_\gamma \left\{
    \int_{\theta_0}^{\theta_0+\pi}\mathbf{W}_\gamma(\theta)d\theta - 
    \int_{\theta_1}^{\theta_1+\pi}\mathbf{W}_\gamma(\theta) d\theta
\right\}.
\end{multline}
It is not hard to see with the help of \eqref{eq:Wcorre_b} that the last
term is $\theta_0-\theta_1$. Therefore,
\begin{equation}
    \frac{\langle\Delta\theta\rangle(\epsilon)}{\pi}= 1-\sum_\gamma
c_\gamma \mathbf{W}_\gamma \left(\frac{\pi}{2}\right).
\end{equation}
Thus the density of states reads ((B.8) in I),
\begin{equation}
    g(\epsilon)=\left|\sum_\gamma c_\gamma
    \frac{d\mathbf{W}_\gamma\left(\frac{\pi}{2}\right)}{d\epsilon}
    \right|
\end{equation}
which can be calculated numerically solving eqs. \eqref{eq:Wcorre_a} for
several values of the energy.
%%%%%%%%%%%%%%%%%%%%%%%%%%%%%%%%%
\subsection{The Lyapunov exponent}
\label{ssec:lya}
The quantity that provides information about the degree of spatial
localization of the electronic states is the
Lyapunov exponent, which gives the exponential growth rate of the
wave function, and thus its inverse is a measure of the localization length
of a state exponentially localized inside the wire. 
The Lyapunov coefficient can be written for the infinite wire as
\begin{equation}
    \Lambda = \lim_{N\rightarrow\infty}\frac{1}{N}\sum_{j=0}^N \log
    \left|\frac{\Psi_{j+1}}{\Psi_j}\right|
\label{eq:lya}
\end{equation}    
where $\Psi_j$ is the amplitude of the wave function at the $j$th atomic site.
Using a mapping technique originally proposed in Ref.\onlinecite{map} we are able to write $\Lambda$ as a
function of the phase (Appendix \ref{ap:lya}),
\begin{equation}
    \Lambda=\frac{1}{2}\lim_{N\rightarrow\infty}\frac{1}{N}\sum_{j=0}^N
    \log F_{\gamma_j}(\theta_j)
\label{eq:lyaF}
\end{equation}
where $F_{\gamma_j}(\theta)=
1-2h_{\gamma_j}(\epsilon)\sin(2\theta)+4h^2_{\gamma_j}(\epsilon)\cos^2\theta$
and $\gamma_j$ stands for the species of the $j$th atom.
Due to the periodicity of $F(\theta)$ we can use the distribution
functions of the phase to calculate the average \eqref{eq:lyaF}:
\begin{equation}
\begin{split}
    \Lambda &=\frac{1}{2} \sum_{\gamma,\beta} c_\gamma p_{\gamma\beta}
    \int_0^\pi \frac{d\mathbf{W}_\gamma(\theta)}{d\theta} \log F_\beta(\theta)
    d\theta =\\
    &= \frac{1}{2} \sum_\gamma c_\gamma \log F_\gamma(\pi) - \frac{1}{2} 
    \sum_{\gamma,\beta} c_\gamma p_{\gamma\beta} \int_0^\pi
    \mathbf{W}_\gamma(\theta) \frac{F'_\beta(\theta)}{F_\beta(\theta)} d\theta.
\end{split}
\end{equation}
From this last expression one can obtain numerically the Lyapunov
exponent as a function of the energy.
%%%%%%%%%%%%%%%%%%%%%%%%%%%%%%%%%%%%%%%%%%%%%%%%%%%%%%%%%%
\section{Results}
\label{sec:results}
\begin{figure}
    \vspace*{-3mm}
    \subfigure{\epsfig{file=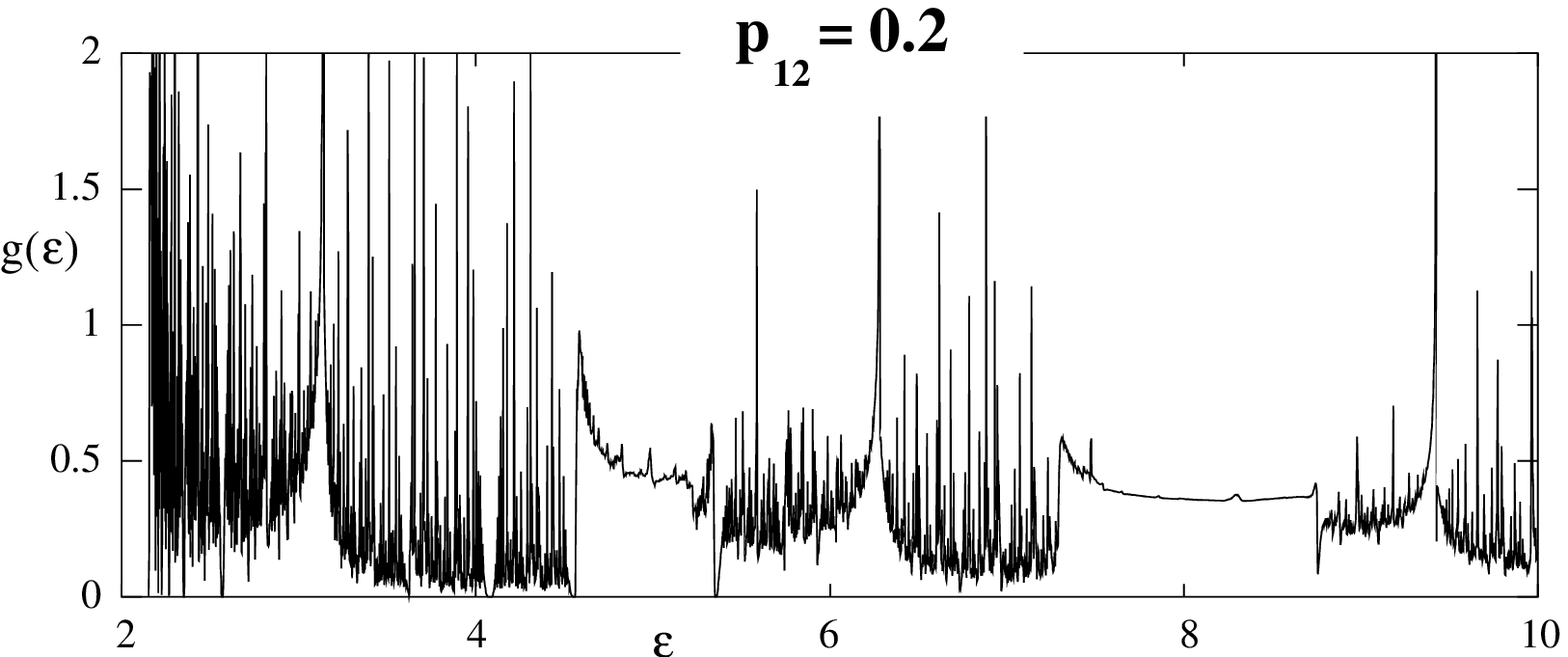,width=\columnwidth}}
    \subfigure{\epsfig{file=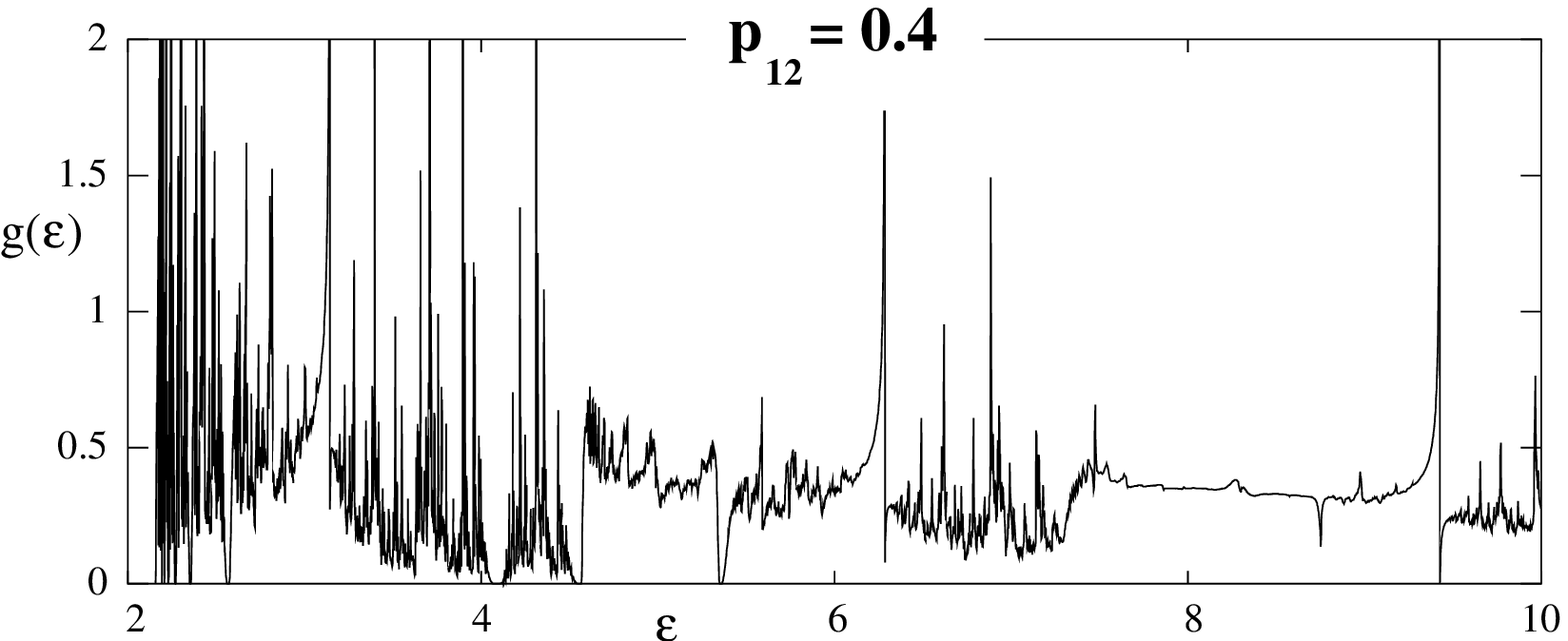,width=\columnwidth}}
    \subfigure{\epsfig{file=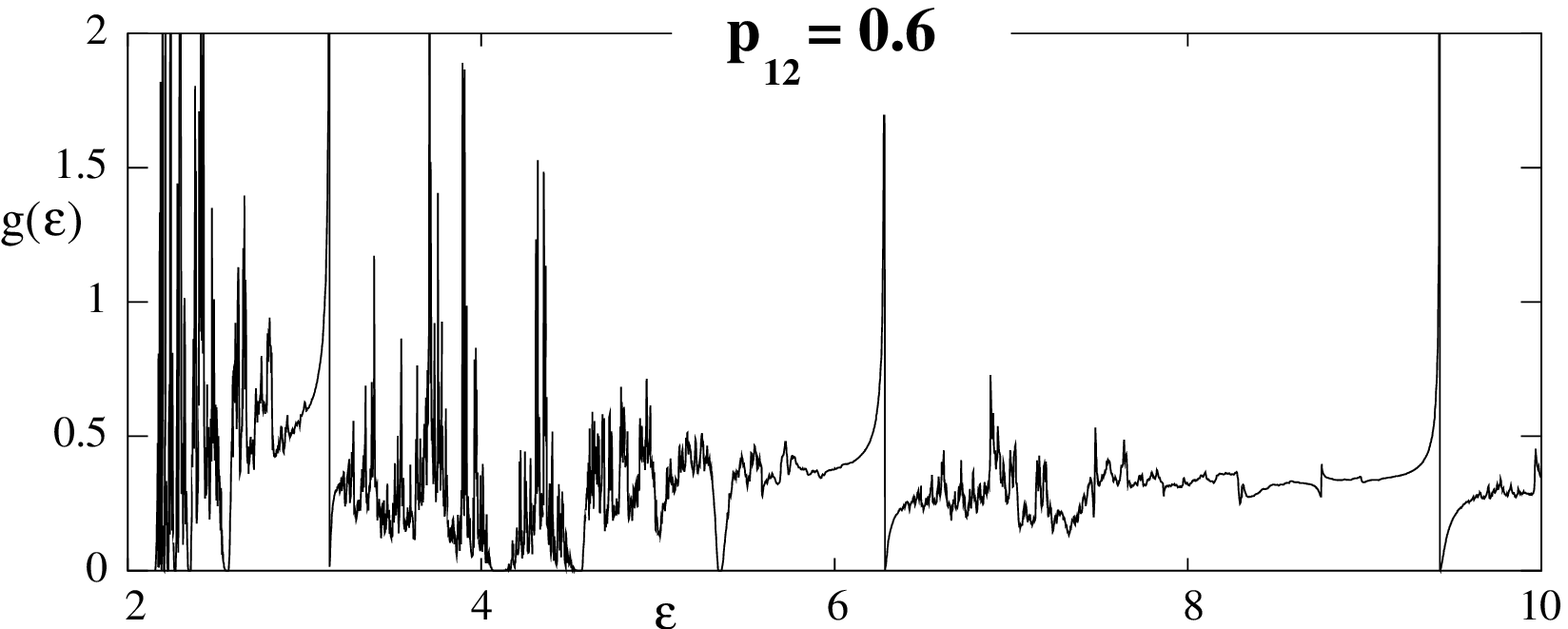,width=\columnwidth}}
    \subfigure{\epsfig{file=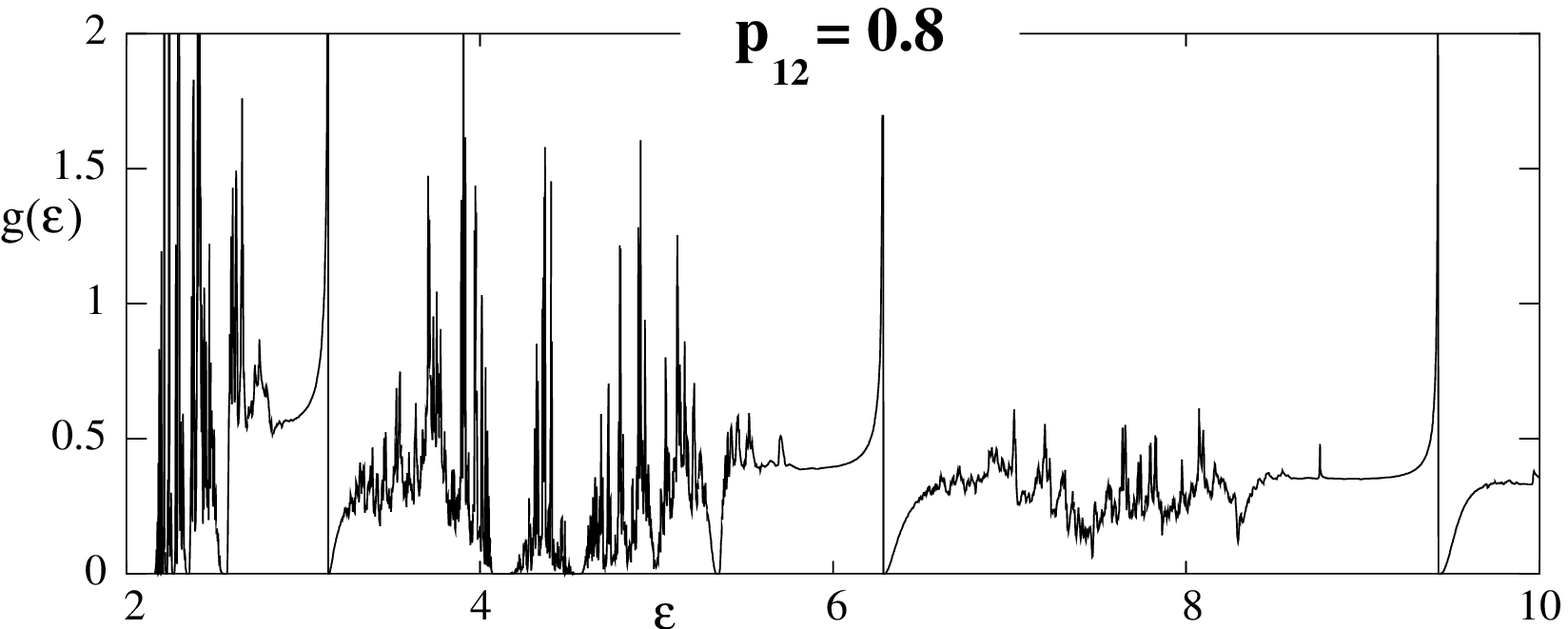,width=\columnwidth}}
    \subfigure{\epsfig{file=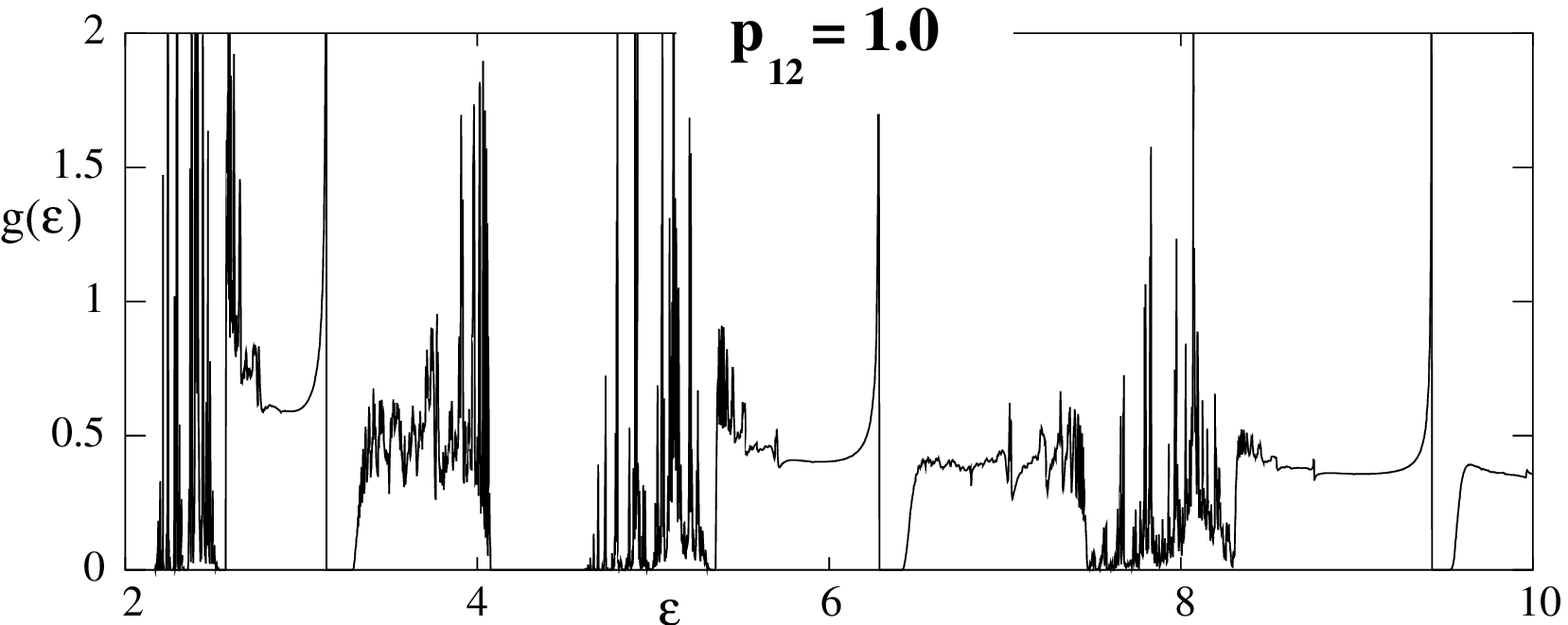,width=\columnwidth}}
    \caption{Density of states for a disordered wire of two
species with different correlations. $\big(\frac{a}{a_1}\big)=-3$ and
$\big(\frac{a}{a_2}\big)=4$ with $c_1=0.4$.}
\label{fig:evolution}
\end{figure}
 Once we have built the mathematical framework lying under the
model, one can try to answer the physical questions that obviously arise
from this new configuration of the disordered quantum wire: how strong is the
effect of this short-range correlations?, does it produce a measurable change
in the density of states or in the localization of the electrons?
 As in Ref. I, we solve the functional equation numerically to obtain the
density of states and to calcultate the Lyapunov exponent. It seems really
hard to find analytical solutions for equations \eqref{eq:Wcorre} if they
exist at all.
\begin{figure*}
    \subfigure{\epsfig{file=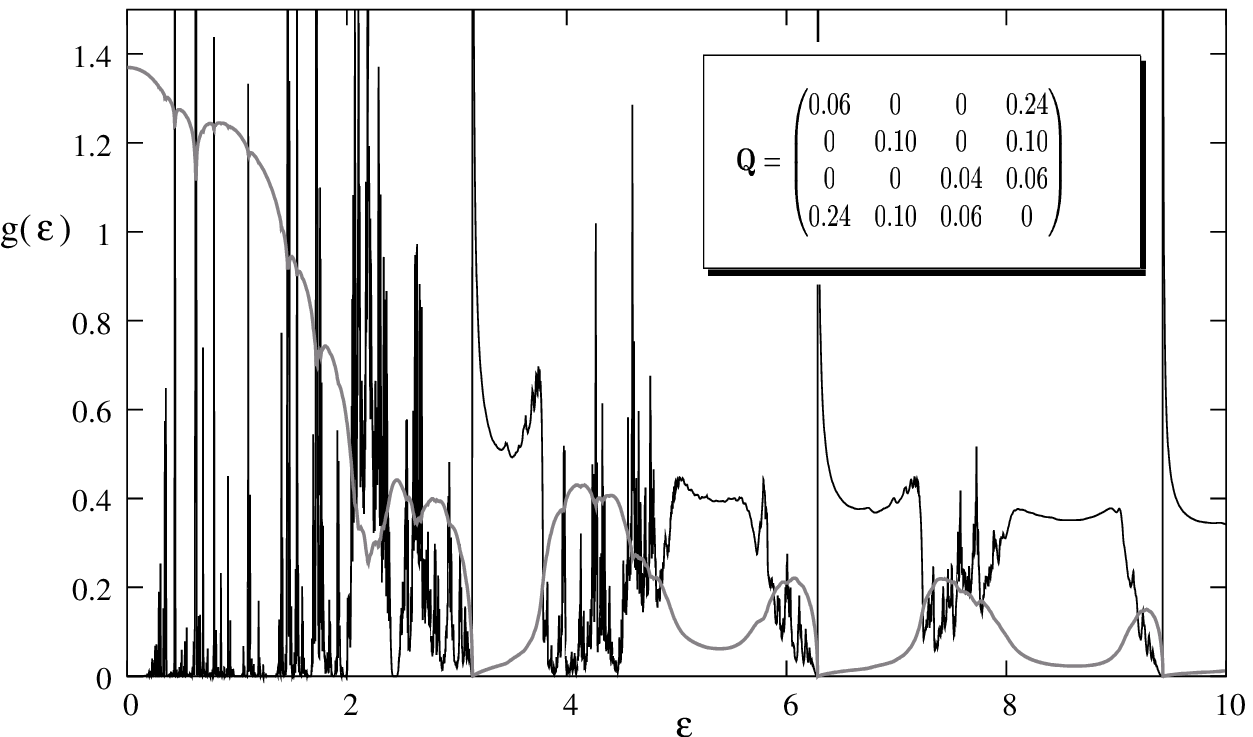,width=\columnwidth}}
    \subfigure{\epsfig{file=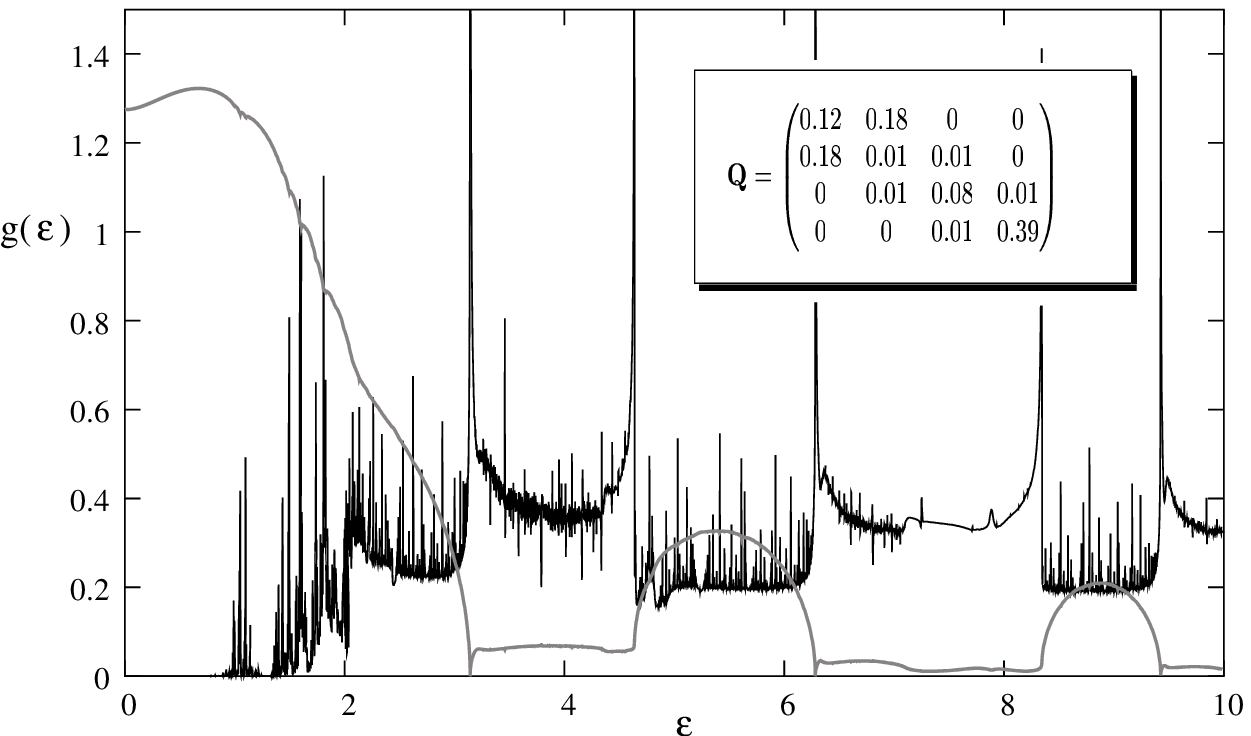,width=\columnwidth}}
    \caption{DOS and Lyapunov exponent (grey line) for a 4 species quantum
wire: $\big(\frac{a}{a_i}\big)= 1, -1, 3, -5$. Notice that the
concentrations are the same for the two configurations}
\label{fig:4species}
\end{figure*}
\begin{figure*}
    \subfigure[$c_1=0.4$]{\epsfig{file=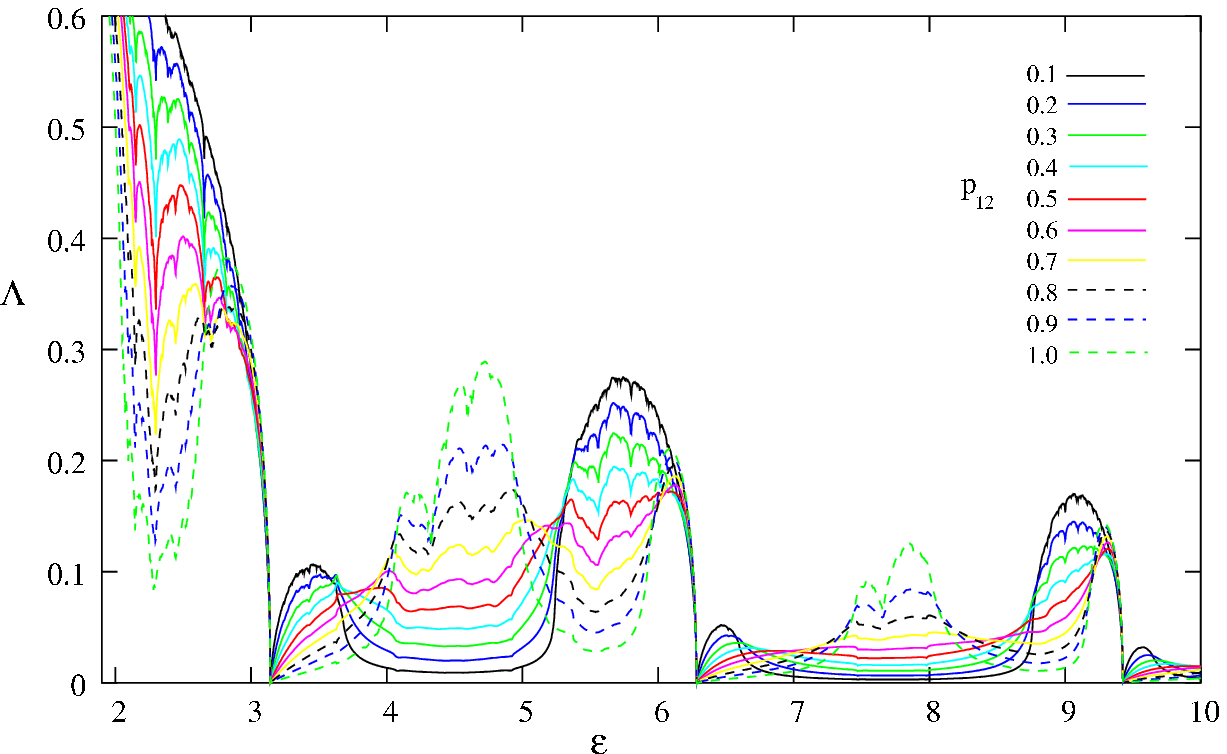,width=\columnwidth}}
    \subfigure[$p_{12}=0.1$]{\epsfig{file=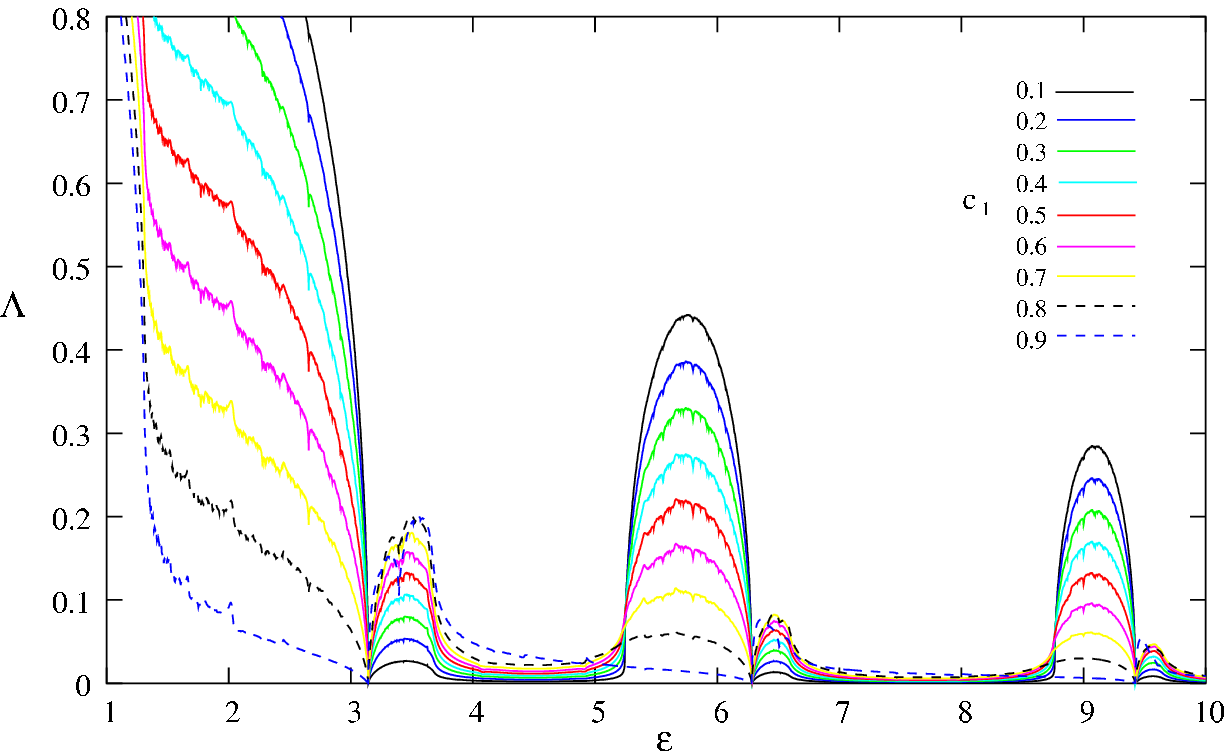,width=\columnwidth}}
\caption{Variation of the Lyapunov exponent for a binary wire composed of
$\big(\frac{a}{a_1}\big)=1$ and $\big(\frac{a}{a_2}\big)=-3$, as a function
of the correlation and as a function of the concentration.}
\label{fig:lya1}
\end{figure*}
Let us analize in the first place the density of states (DOS) for our quantum wires.
In Figure \ref{fig:evolution} the evolution of the DOS for a binary wire as
a function of the correlation $p_{12}$ is shown for a certain value of the
concentrations. This distribution of states is drastically changed from the
initial situation in which the probability to find the cluster -12- is low
to the final stage when we impose the atoms of species 1 to appear always
isolated. Note however that the concentrations are the same in both cases!. Another
example can be seen in Figure \ref{fig:4species} for a wire composed of 4
species. As the number of species increases, the number of correlations
grows as $\frac{m(m-1)}{2}$, and the distribution of states can adopt a lot
of different shapes: the exploration of the whole correlations and
concentrations space
for $m>2$ can take a long time.
From these graphics we conclude that the
correlations can unbalance the spectra of the disorder system quite far from the
completely random configuration. In fact the correlation can be tuned to
open (close) an energy gap or to increase (decrease) the number
of available states in a certain energy interval without changing the
concentrations of the atomic components. The variation of the DOS with the correlations is of course
a function of the concentrations being the chain with
homogenized participation the one whose distribution has the strongest
dependence on $\mathbf{P}$. Let us also remark that the fractal behaviour
of the density of states in certain energy ranges, reported in reference I,
still manifests itself for the different
correlations regimes.

\begin{figure*}[t]
    \subfigure[$c_1=0.5$]{\epsfig{file=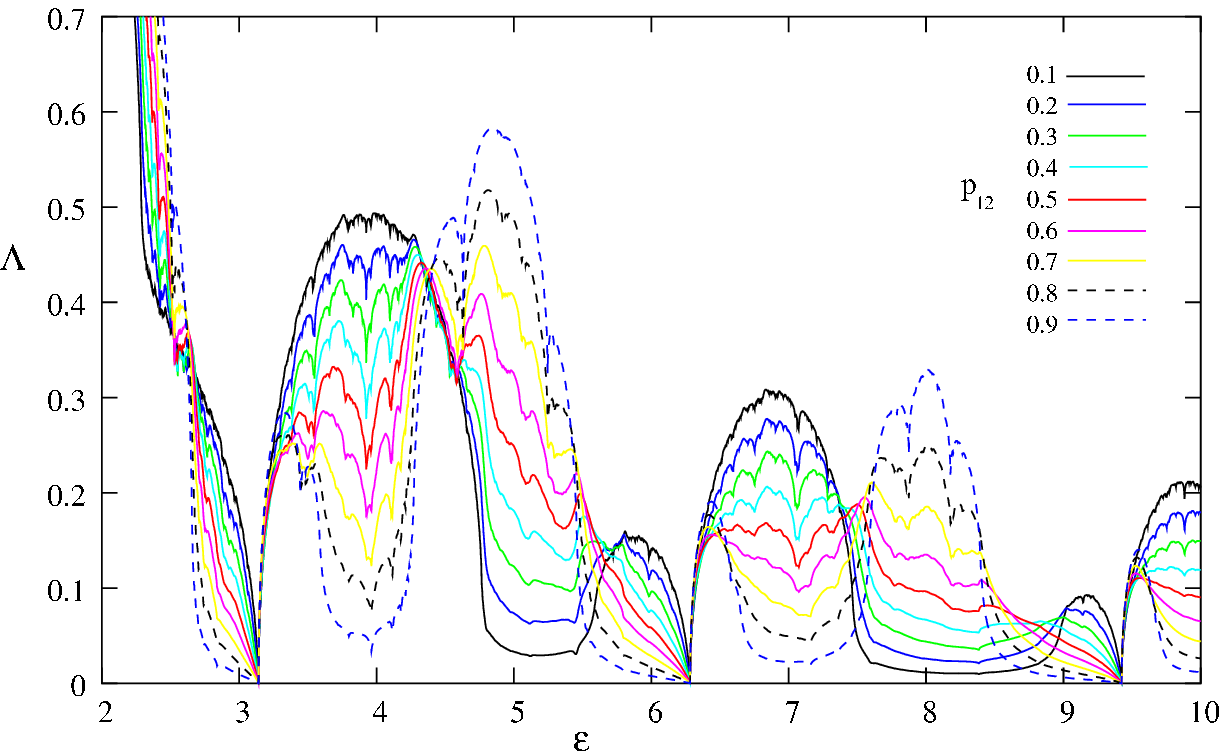,width=\columnwidth}}
    \subfigure[$p_{12}=0.5$]{\epsfig{file=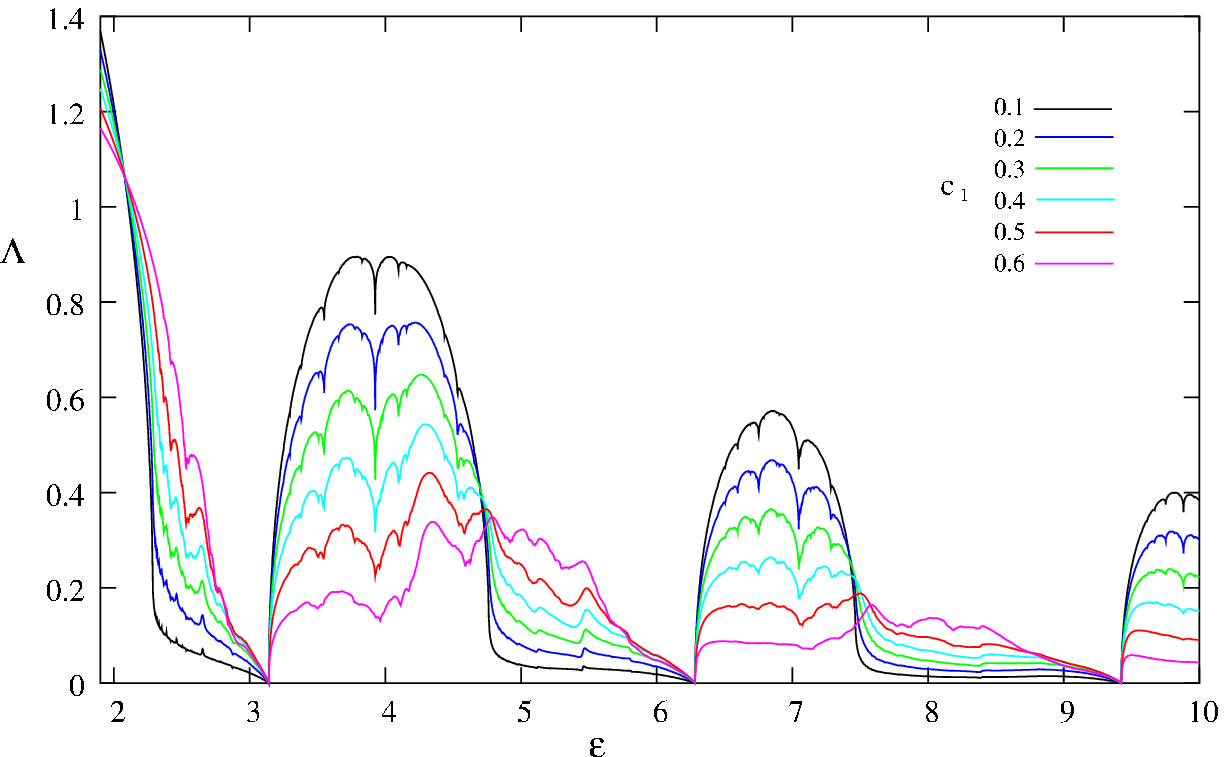,width=\columnwidth}}
\caption{Variation of the Lyapunov exponent for a binary wire composed of
$\big(\frac{a}{a_1}\big)=-2$ and $\big(\frac{a}{a_2}\big)=5$, as a function
of the correlation and as a function of the concentration.}
\label{fig:lya2}
\end{figure*}
Let us have a look at the localization of the electronic states. 
 In Figures \ref{fig:lya1} and \ref{fig:lya2} the behaviour of the
Lyapunov exponent for a binary chain changing the correlations and the
concentrations is shown. In Figure \ref{fig:4species} the different shapes
of the degree of localization for different correlations can also be
noticed for a 4-species wire. 
The effective influence of the correlations
on the localization of the electrons inside the quantum wires is
established. In the study of correlated disorder two main kinds of
correlations have been taken into account namely short-range and
long-range. The former, mainly represented by the random dimer model
\cite{RD}, has been widely accepted to be the reason for  the emergence of resonant energies
and a set of states close to the resonant one for which the localization
length becomes larger that the system size \cite{kp-adame}, improving greatly the transport
properties \cite{bellani}. The long-range correlations seem to be
able to include in the spectrum mobility edges \cite{Moura},  that can be
entirely controlled by the correlators  \cite{Izkro}, yielding an energy
interval of a complete transparency.
In our type of wires we have not observed mobility edges nor resonant energies.
 However an important effect on the localization of the states is shown that
seems to act globaly in the whole energy range in contrast to the more
restricted effect of others short-range correlated disorder models. In fact
for certain energies the Lyapunov exponent can be decreased an 80-90\% of
its maximun value changing the correlations at the expense of an
increasing behaviour for other energies. Whether these correlations 
might cause $\Lambda$ to go below the inverse of the length of
a finite sample for some energy intervals is not certainly clear but it could
be a point for future discussions. One can now argue if this
influence of the correlations on the localization as well as on the DOS
will be a measurable effect on short finite wires that represent a simple 
practical implementation, or on the contrary is just theoretically
achievable only in  infinite
unreal systems. In order to answer that question and ascertain ourselves of
the importance of the results we learn to generate finite sequences showing
this kind of correlations. 
\begin{figure}
    \epsfig{file=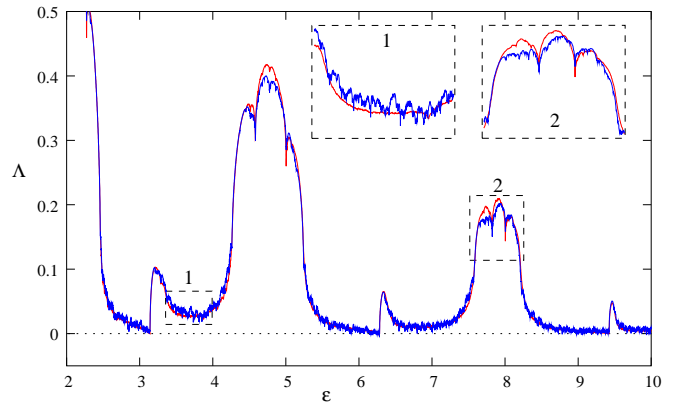,width=\columnwidth}
\caption{Lyapunov exponent for a binary wire with
$\big(\frac{a}{a_1}\big)=-2$, $\big(\frac{a}{a_2}\big)=3$, $c_1=0.5$
and  $p_{12}=0.9$. The red line corresponds to the infinite wire. The blue
line shown $\Lambda$ numerically calculated for a sequence of 1000 atoms.}
\label{fig:finite}
\end{figure}
In Figure \ref{fig:finite} we show the Lyapunov exponent of an infinite
wire for a certain concentration and correlation calculated using the
functional equation, and the one obtained numerically from equation
\eqref{eq:lya} for a correlated disordered chain of 1000 atoms. The
convergence is excellent with only one realization of the correlated
disorder using one thousand of atoms, and we have also
checked it for others couplings of the species and other correlations
obtaining a high degree of agreement. Thus it means that the described
results are real effects with some practical importance on experimental
realizations and on the other hand that the method of the functional
equation, although  designed for an infinite wire, can reproduce quite well
the tendency in finite samples. 
%%%%%%%%%%%%%%%%%%%%%%%%%%%%%%%%%%%%%%%%%%%%%%%%%%%%%%%%%%%%%%%%%%%%%%%%%%%%%%%%
\section{Concluding remarks}
    To summarize, we have extended the model proposed in reference I and
completed the formalism introduced there including the possible
correlations which must be present in any realistic model of one
dimensional conductance. The method described here is the most natural manner
to account for correlations that can manifest in nature or even inside a
manufactured quantum wire in a nonintentional way. 

It has been established how due to the correlated
disorder the density of states is essentially modified and the localization
of the electronic states changes in a non negligible proportion which might
alter some macroscopic properties of these structures. 

In addition to the obtained results we are preparing a future report in
which we will construct the functional equation formalism in a more
coherent way relating it to others methods currently used. There it will be  shown
how this mathematical framework is flexible enough to be suitable for
applying it to different disorder models such as the Tight-Binding model as well as
other correlation schemes such as the random dimer.

%%%%%%%%%%%%%%%%%%%%%%%%%%%%%%%%%%%%%%%%%%%%%%%%%%%%%%%%%%%%%%%%%%%%%%%%%%%%%%%
\begin{acknowledgments}
    We acknowledge with thanks the support provided by the Research in
Science and Technology Agency of the Spanish Goverment (\textbf{DGICYT})
under contract \textbf{BFM2002-02609}.
\end{acknowledgments}

%%%%%%%%%%%%%%%%%%%%%%%%%%%%%%%%%%%%%%%%%%%%%%%%%%%%%%%%%%%%%%%%%%%%%%%%%%%%%%%%
\appendix
\section{Lyapunov exponent}
\label{ap:lya}
It can be shown that the  electronic wave function satisfies,
\begin{equation}
    2h_{\gamma_j}(\epsilon)\Psi_j=\Psi_{j+1}+\Psi_{j-1}
\label{eq:canonical}
\end{equation}
where $\Psi_j$ is the amplitude of the wave function at the $j$th atomic
site of the chain. This equation can be visualized as a two dimensional
mapping in polar coordinates with the identification
$\Psi_{j+1}=r_{j+1}\cos\theta_{j+1}$ and $\Psi_j=r_{j+1}\sin\theta_{j+1}$,
\begin{equation}
    \begin{pmatrix} r_{j+1}\cos\theta_{j+1} \\
    r_{j+1}\sin\theta_{j+1}\end{pmatrix}=\begin{pmatrix}
    2h_{\gamma_j}(\epsilon) & -1 \\ 1 & 0 \end{pmatrix}
    \begin{pmatrix} r_j\cos\theta_j \\
    r_j\sin\theta_j\end{pmatrix}
\end{equation}
which yields the known transmission relationship for the phase and the
following for the moduli $r$,
\begin{equation}
 \frac{r_{j+1}^2}{r_j^2}\equiv F_{\gamma_j}(\theta_j)=
  1-2h_{\gamma_j}(\epsilon)\sin(2\theta_j)+4h^2_{\gamma_j}(\epsilon)\cos^2\theta_j.  
\label{eq:moduli}
\end{equation}
If now we introduce the polar coordinates into the definition of the
Lyapunov exponent we obtain:
\begin{multline}
    \Lambda = \lim_{N\rightarrow\infty}\frac{1}{N}\sum_{j=0}^N \log
\left|\frac{\Psi_{j+1}}{\Psi_j}\right|\\=\lim_{N\rightarrow\infty}\frac{1}{N}\sum_{j=0}^N \log
\left(\frac{r_{j+1}}{r_j}\right)+ \lim_{N\rightarrow\infty}\frac{1}{N}\sum_{j=0}^N \log
\left|\frac{\cos\theta_{j+1}}{\cos\theta_j}\right|\\=\lim_{N\rightarrow\infty}\frac{1}{N}\sum_{j=0}^N \log
\left(\frac{r_{j+1}}{r_j}\right)
\end{multline}
because according to the meaning of $\Lambda$ the cosinus term must be zero.
And using \eqref{eq:moduli} we can calculate the Lyapunov coefficient as a
function of the phase, as:
\begin{equation}
    \Lambda=\frac{1}{2}\lim_{N\rightarrow\infty}\frac{1}{N}\sum_{j=0}^N
\log F_{\gamma_j}(\theta_j).
\end{equation}
which is our expression \eqref{eq:lyaF}.
%%%%%%%%%%%%%%%%%%%%%%%%%%%%%%%%%%%%%%%%%%%%%%%%%%%%%%%%%%%%%%%%%%%%%%%%%%%%%


\begin{thebibliography}{}
\bibitem{QWI}
    J.M. Cerver\'o, A. Rodr\'{\i}guez, Eur. Phys. J B \textbf{30}, 239 (2002)
\bibitem{map}
    F.M. Izrailev, T. Kottos, G.P. Tsironis, Phys. Rev. B \textbf{52}, 3274
(1995)
\bibitem{RD}
    D.H. Dunlap, H-L. Wu, P.W. Phillips, Phys. Rev. Lett. \textbf{65}, 88
(1990)
\bibitem{kp-adame}
    A. S\'anchez, E. Maci\'a, F. Dom\'{\i}nguez-Adame,
Phys. Rev. B. \textbf{49}, 147 (1994)
\bibitem{bellani}
    V. Bellani, E. Diez, R. Hey, L. Tony, L. Tarricone, G.B. Parravicini,
F. Dom\'{\i}nguez-Adame, R. G\'omez-Alcal\'a, Phys. Rev. Lett. \textbf{82},
2159 (1999)
\bibitem{Moura}
    F.A.B.F. de Moura, M.L. Lyra, Phys. Rev. Lett. \textbf{81}, 3735 (1998)
\bibitem{Izkro}
    F.M. Izrailev, A.A. Krokhin, Phys. Rev. Lett. \textbf{82}, 4062 (1999)
\end{thebibliography}
\end{document}